\documentclass[aps,prb,twocolumn,psfig,showpacs]{revtex4}
\usepackage{amsfonts}
\usepackage{amsmath}
\usepackage{graphicx}
\usepackage{bm}
\usepackage{amssymb}
\usepackage{times}
\usepackage{dcolumn}

\begin{document}

%\draft

\title{Matrix Product State and Quantum Phase Transitions in the One-Dimensional Extended Quantum Compass Model}

\author{Guang-Hua Liu$^{1}$, Wei Li$^{2}$}

\email[Corresponding author. ]{Email: liwei-b09@mails.gucas.ac.cn}

\author{Wen-Long You$^{3}$}
\author{Guang-Shan Tian$^{4}$}
\author{Gang Su$^{2}$}
\address{1. Department of Physics, Tianjin Polytechnic University, Tianjin 300387,China\\
2. Theoretical Condensed Matter Physics and Computational Materials Physics Laboratory, College of Physical Sciences,Graduate University of Chinese Academy of Sciences, P. O. Box 4588, Beijing 100049, China\\
3. School of Physical Science and Technology, Soochow University, Suzhou, Jiangsu 215006, China\\
4. School of Physics, Peking University, Beijing 100871, China}

%\date{\today}

\begin{abstract}

The matrix product state (MPS) is utilized to study the ground-state
properties and quantum phase transitions (QPTs) of the
one-dimensional extended quantum compass model (EQCM). The MPS wavefunctions
are argued to be very efficient descriptions of the ground states,
and are numerically determined by imaginary time projections. The
ground-state energy, correlations, quantum entanglement and its
spectrum, local and nonlocal order parameters, etc., are calculated
and studied in details. It is revealed that the von Neumann
entanglement entropy, as well as the nearest neighbor correlation
functions, can be used to detect the second-order QPTs, but not the
first-order ones, while fidelity detections can recognize both. The
entanglement spectrum is extracted from the MPS wavefunction, and
found to be doubly degenerate in disordered phases, where
nonzero string order parameters exist. Moreover, with linearized
tensor renormalization group method, the specific heat curves are
evaluated and their low temperature behaviors are investigated.
Compared with the exact solutions, our results verify
that these MPS-based numerical methods are very accurate and
powerful, and can be employed to investigate other
EQCMs which do not permit exact solutions at present.

\end{abstract}


\maketitle

%\narrowtext

\newpage

\section*{I. Introduction}

During the past several decades, the role of the orbital degrees of freedom in determining the magnetic and transport properties of transitional-metal oxides (TMOs) has been widely recognized.\cite{Goodenough,Tokura,Aken,Hotta,Zenia,Wakabayashi} The complex intrinsic interplay in TMOs induces their extremely rich phase diagrams and various fascinating physical phenomena. In order to mimic these orbital states with a two-fold degeneracy, the quantum compass model (QCM) was firstly introduced by Kugel and Khomskii.\cite{Kugel} The orbital degrees of freedom are represented by pseudospin-1/2 operators, and the competition between orbital orderings in different directions is simulated by anisotropic couplings between these pseudospins. Particularly, two-dimensional (2D) QCM has attracted considerable attention due to its interdisciplinary character. Besides the ability to describe $t_{2g}$ systems, it was also proposed that the compass model can describe the physics of protected qubits,\cite{Doucot,Milman} and hence it may have potential application in the quantum information techniques. The strong quantum frustration makes it difficult to solve the system analytically, and consequently leads to large degeneracy in the energy spectrum, which sets obstacles for numerical simulations. \cite{Wenzel} It is generally implied that there exist a symmetry-broken ground state and a first-order quantum phase transition (QPT) at the self-dual point. \cite{Dorier,Chen,orus-2dqcm,JVidal,You2011}

On the other hand, the one-dimensional (1D) QCM has also triggered extensive studies.\cite{Brzezicki,Brzezicki2,Sun1,Sun2,Jafari,You,Eriksson,Mahdavifar} In Ref.\onlinecite{Brzezicki}, by mapping to the quantum Ising model, Brzezicki \emph{et al.} obtained an exact solution of the 1D extended QCM (EQCM), revealing that it exhibits a first-order transition between two disordered phases. Subsequently, Wen-Long You and Guang-Shan Tian adopted the reflection positivity technique in the standard pseudospin representation to rigorously determine the ground-state degeneracy. \cite{You} And, a first-order phase transition was also confirmed. Following the approach in Ref.\onlinecite{Brzezicki},  Eriksson and Johannesson \cite{Eriksson} studied the QPTs in a 1D EQCM with more tunable parameters. They suggested that the reported first-order phase transition in fact occurs at a multicritical point where a line of the first-order transition meets with a line of the second-order transition. Generally speaking, a first-order QPT is often associated with energy level crossing in the ground state, and hence the entanglement measures, such as concurrence and entanglement entropy, would behave discontinuously.\cite{Gu,Liu} However, in Ref. \onlinecite{Eriksson}, the authors claimed that they encountered an ``accidental" exception. The concurrence and block entanglement can accurately signal the second-order transitions, but not the first-order ones. In other words, the entanglement measures do not show any discontinuities or singularities across the first-order quantum critical points (QCPs) in the 1D EQCM. Nevertheless, a converse point of view that both concurrence and quantum discord can reliably detect the first-order QPTs of this model was proposed very recently.\cite{You2} In a sense the 1D EQCM can be exactly solved by taking Jordan-Wigner transformation, nevertheless, it is still not easy to analytically calculate the spin correlations for arbitrary sites and the excited states.

In this paper, we investigate numerically the ground-state properties and QPTs
of 1D EQCM with matrix product state (MPS) variational wavefunction and the related algorithms.
We would like to point out that MPS is a very useful and highly efficient
real-space description of the ground states, and it provides a novel way to
study the QPTs in EQCM. To be specific, firstly, some exact MPS ground states
for EQCM Hamiltonian in some limiting cases can be obtained, and for off-limiting generic parameters, infinite time-evolving block decimation (iTEBD) algorithm \cite{Vidal} is adopted to determine the variational MPS ground state. Very accurate results can be achieved in
gapped regions (up to 8$\sim$9 digits compared with exact solution, see Fig. \ref{fig02} below)
with a small number of reserved states. In addition, the iTEBD method can also be employed to take adiabatic continuation calculations, which apparently reveals the energy level crossing around first-order QPTs. Secondly, given the real space wave-function in MPS form, the interesting quantities including the ground-state energy, energy spectrum, correlation functions, entanglement entropy, fidelity per site, as well as local and nonlocal order parameters, \emph{etc}, can be conveniently evaluated. Some of them are not easy to be obtained by other methods. Thirdly, the MPS-based algorithms can be applied to other extended models, and hence provide us powerful tools to explore other EQCMs without exact solutions.

Through the numerical calculations with MPS, we verify the phase diagram of the 1D EQCM (see Eq. (\ref{Hamiltonian})), and it is uncovered that both the first- and second-order QPTs can be detected by the fidelity, while the entanglement measures can only capture the later ones. Furthermore, we discover that the entanglement spectra in disordered phases of EQCM happens to be doubly degenerate, and correspondingly there exist two non-local string order parameters, which reveals the hidden $Z_2 \times Z_2$ symmetry breaking.

This paper is organized as follows. In section II, the Hamiltonian of the
1D EQCM is introduced, along with the MPS description and related
perturbation analysis. Besides, the entanglement and fidelity measures
in the framework of MPS are concerned and discussed. In sections III, we provide our main numerical results, which include the ground-state energy, entanglement entropy, fidelity and string order parameters in different regions of the phase diagram. Afterwards in section IV, with finite temperature algorithm, i.e., linearized tensor renormalization group (LTRG), the specific heat curves of 1D EQCM are calculated and analyzed. Finally, some possible extensions of present work, as well as a summary, are presented in section V.

\section*{II. Model Hamiltonian, Matrix Product State, Entanglement Entropy and Fidelity}

\subsection{Quantum Compass Model}
The 1D EQCM is given by
\begin{equation}
\hat{H} = \sum^{N'}_{i=1} (J_1 \sigma^{z}_{2i-1} \sigma^{z}_{2i} + J_2 \sigma^{x}_{2i-1} \sigma^{x}_{2i} + L_1 \sigma^{z}_{2i} \sigma^{z}_{2i+1}),
\label{Hamiltonian}
\end{equation}
where periodic boundary condition is assumed, and $N=2N'$ is the total number of sites.
The $\sigma_{i}^{x,z}$ are Pauli matrices on the $i$th site, $J_1$, $J_2$ on odd bonds, along with $L_1$ on even bonds, are exchange couplings. For the following calculations in sections III and IV, the coupling constant $L_1=1$ in Hamiltonian Eq. (\ref{Hamiltonian}) is set as energy scale.

\begin{figure}
\begin{center}
\includegraphics[width=1.0\linewidth]{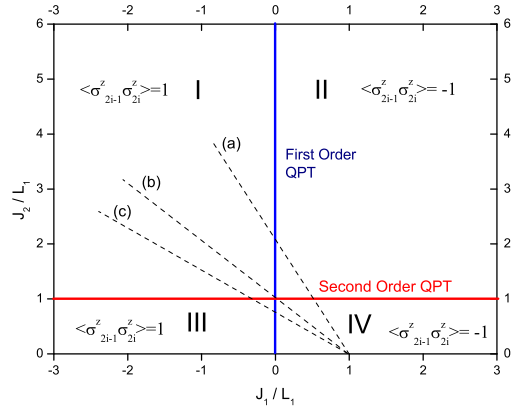}
\end{center}
\caption{(Color online) Schematic phase diagram of the 1D EQCM, the four different phases are marked as regions I, II, III, and IV. The dashed lines denote three typical paths ((a) $J_2 = 2(1-J_1)$; (a) $J_2 = 1-J_1$; (c) $J_2 = 0.8(1-J_1)$) that will be used in the following discussions. $L_1=1$ is set as the energy scale.}
\label{Fig1}
\end{figure}

The ground-state phase diagram of 1D EQCM (see Fig. \ref{Fig1}) is sketched by previous studies.\cite{Eriksson} As is shown in  Fig. \ref{Fig1}, the system undergoes a first-order QPT identified by critical line $J_1 = 0$ and a second-order QPT with critical line $J_2/L_1 = 1$. A multicritical point ($J_1 = 0, J_2/L_1 =1$) locates where the lines of the first-order and second-order QPTs meet.

\begin{figure}
\begin{center}
\includegraphics[width=0.9\linewidth]{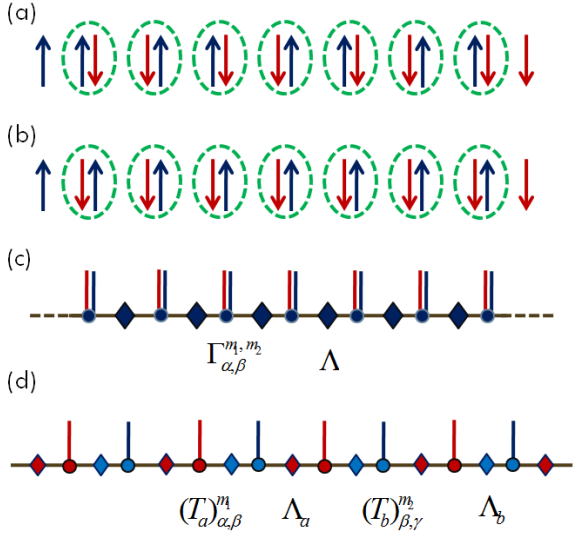}
\end{center}
\caption{(Color online) (a) and (b) show two typical spin configurations, the spins on odd bonds are parallel (for $J_1<0$) or anti-parallel (for $J_1>0$). The dash ovals denote the spins on even bonds. (c) and (d) show the one- ($\Gamma$) and two-period ($T_a$, $T_b$) MPS wavefunctions, $\Lambda$ is a diagonal matrix on each bond.}
\label{fig01}
\end{figure}

The Hamiltonian (\ref{Hamiltonian}) commutes with the parity operators $P_i \equiv \sigma^{x}_{2i-1} \sigma^{x}_{2i}$, and thus the parity of every odd bond is conserved. In such circumstance, the Hilbert space can then be decomposed into subspace $V ({p_i})$, where $p_i$ is the eigenvalue of $P_i$ and introduced to label the relative pseudospin direction on odd bonds, that is, $p_i=0$ ($p_i=1$) when the two pseudospins are parallel (antiparallel). It is disclosed that the ground state lies in space $\{ p_1 = p_2 = ... = p_{N'} = 0\}$ for $J_1<0$, and in $\{ p_1 = p_2 = ... = p_{N'} = 1\}$ for $J_1>0$,\cite{You} that is, for ferromagnetic coupling $J_1$, two spins on odd bonds can only be parallel (one of such spin configurations is shown in Fig. \ref{fig01} (a)), while for antiferromagnetic (AF) coupling $J_1$, the spins on odd bonds must be antiparallel in the ground state (see Fig. \ref{fig01} (b)).

\begin{figure}
\begin{center}
\includegraphics[width=0.9\linewidth]{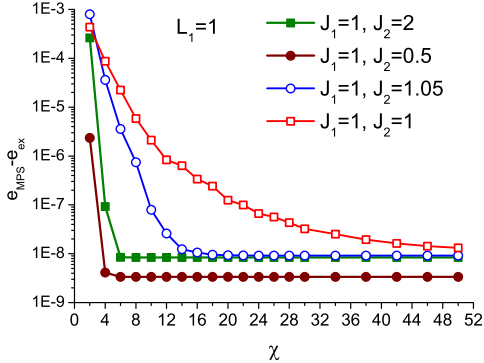}
\end{center}
\caption{(Color online) The calculation errors of energy per site, $e_{\rm{MPS}}$ means MPS numerical result and $e_{\rm{ex}}$ is the exact solution. For non-critical regions, even with smallest bond dimensions ($\chi$ =  2 or 4), the MPS calculation provides very accurate results. The systems at or near the critical line ($J_2/J_1=1.0$ and $1.05$ shown above) are harder to tackle, however, a few more states (say, $\chi = 20$ or 30) is adequate in practical calculations.}
\label{fig02}
\end{figure}

\subsection{Matrix Product States and Perturbation Analyses}

In principle, any quantum state of many-body system can be expressed by MPS form through taking successive Schmidt decompositions site by site.\cite{Schmidt, Cirac} However, not all of these MPS expressions are efficient and can be utilized for simulations. For 1D cases, thanks to the entanglement entropy area law,\cite{Eisert} quantum spin chain with only finite range interactions and possesses a gapped spectrum can be efficiently simulated with the MPS-based algorithms. MPS is closely related with the density matrix renormalization group method,\cite{White} and it well satisfies the entanglement area law in 1D. When the the chain is divided into two blocks by cutting an bond, the renormalized left (right) bases are the eigenvectors of the reduced density matrix subsystem to the left(right) of the broken bond. As long as the entanglement entropy between two subsystems is bounded by the area law, the classical simulation of the 1D quantum system can be performed efficiently.

For the present 1D EQCM, as shown in Fig. \ref{fig01} (c), a tensor $\Gamma^{m_{2i},m_{2i+1}}$ is used to address the two spins on even bonds, where $m$ means local spin physical index, and the wavefunction in uniform (period-one) MPS form can be written as
\begin{eqnarray}
| \Psi_{\textrm{MPS}}^{\rm{I}} \rangle = \textrm{Tr}( \prod_{i=1}^{N'} \Gamma^{m_{2i}, m_{2i+1}} \Lambda_i) | ..., (m_{2i}, m_{2i+1}), ... \rangle,
\label{MPS}
\end{eqnarray}
in which, $\Lambda$ means a $\chi \times \chi$ diagonal matrix, $\chi$ is also called the bond dimension, and $\textrm{Tr}$ is the trace of matrix product.

To explain this point more explicitly, we take the limiting cases with $L_1=0$ into account, where the spins on even bonds are unentangled, and the exact MPS ground states are thus obtainable. For $J_1<0$ (and $J_2>0$), the limit parameter point belongs to region I of the phase diagram Fig. \ref{Fig1}. The ground state of the local two-site Hamiltonian on odd bond is
\begin{equation}
| \phi_{f}(2i-1,2i) \rangle =  \frac{1}{\sqrt{2}} | \uparrow_{2i-1} \uparrow_{2i} - \downarrow_{2i-1}  \downarrow_{2i} \rangle,
\label{eq-wavefunc-odd-bond}
\end{equation}
with bond energy $J_1-J_2$, and it requires that the spin orientations on odd bonds must be parallel. Thus MPS Eq. (\ref{MPS}) with the following projection tensor $\Gamma$ (bond dimension $\chi=2$) is the true ground state of the system: $\Gamma^{\uparrow\uparrow}_{1,1} = \Gamma^{\uparrow\downarrow}_{1,2} = 1$, and $\Gamma^{\downarrow\uparrow}_{2,1} = \Gamma^{\downarrow\downarrow}_{2,2} = -1$ (the negative sign originates from the minus sign between spin up and down components in Eq. (\ref{eq-wavefunc-odd-bond})); similarly, for $J_1>0$, the limit case locates at region II, and the ground state on odd bond is
\begin{equation}
| \phi_{af}(2i-1,2i) \rangle = \frac{1}{\sqrt{2}} | \uparrow_{2i-1} \downarrow_{2i} - \downarrow_{2i-1} \uparrow_{2i} \rangle,
\end{equation}
with energy $-J_1-J_2$, and the tensor $\Gamma$ in ground-state MPS
is as: $\Gamma^{\downarrow\uparrow}_{1,1} =
\Gamma^{\downarrow\downarrow}_{1,2} = 1$, and
$\Gamma^{\uparrow\uparrow}_{2,1} = \Gamma^{\uparrow\downarrow}_{2,2}
= -1$. In addition, for both cases, $\Lambda = \frac{1}{\sqrt{2}}I$
is a $2\times2$ diagonal matrix with doubly degenerate values.

In the practical iTEBD projection process, the MPS wavefunction is usually organized as two-period, i.e., it consists of two types of tensors $T_a$ and $T_b$,
\begin{equation}
|\Psi_{\rm MPS}^{\rm{II}} \rangle = {\rm Tr}(\prod_{i}^{N'} \Lambda_a T_a^{m_{2i-1}} \Lambda_b T_b^{m_{2i}}) |..., m_{2i-1}, m_{2i}, ...\rangle,
\end{equation}
where $\Lambda_a$ ($\Lambda_b$), is $\chi_{a(b)} \times \chi_{a(b)}$ diagonal matrix on the corresponding bond. The exact MPS expressed with one tensor $\Gamma$ can also be rewritten with $T_a$ and $T_b$. Because the spins on even bonds are unentangled, bond dimension $\chi_a = 1$ and $\Lambda_a = 1$, the other bond dimension $\chi_b = 2$, with diagonal matrix $\Lambda_b=\frac{1}{\sqrt{2}}I$. When $J_1<0$, the nonzero tensor elements are $(T_b)_{1,1}^{\uparrow} = 1$, $(T_b)_{2,1}^{\downarrow} = -1$, and  $(T_a)_{1,1}^{\uparrow}=(T_a)_{1,2}^{\downarrow}=1$; while for $J_1>0$, $(T_b)_{1,1}^{\downarrow} = 1$, $(T_b)_{2,1}^{\uparrow} = -1$, and again $(T_a)_{1,1}^{\uparrow}=(T_a)_{1,2}^{\downarrow}=1$.

Besides regions I and II, there also exist exact MPS ground states in regions III and IV. In Fig. \ref{Fig1}, those parameter points are along the line $J_2=0$. Owing to the absence of quantum fluctuations, the model reduces to a classical Ising model with alternating couplings $J_1$ and $L_1$, and the ground state is direct product state, i.e., MPS with bond dimension $\chi=1$. When $J_1>0$, one ground state spin configuration is illustrated in Fig. \ref{fig01} (b), and the MPS is two-period, with $(T_a)^{\uparrow}_{1,1}=1$ and $(T_b)^{\downarrow}_{1,1}=1$. While for $J_1<0$, one spin configuration is shown in Fig. \ref{fig01} (a), and the MPS is four-period, i.e., consists of four $T$ tensors. Ignoring the bond indices owing to $\chi=1$, the nonzero elements are $T_{4n-3}^{\uparrow} = T_{4n-2}^{\uparrow} = 1$, and $T_{4n-1}^{\downarrow} = T_{4n}^{\downarrow} = 1$, where $4n-i$ is used to mark the lattice site, and $n=1,2,...,N'/2$ ($N'$ is assumed to be even number).

Apart from the above limiting points, the MPS ground state can not be written down generally, however, we can adopt the ordinary perturbation theory to argue that MPS description is still a very nice ground state approximation. Firstly we take regions I and II as examples, and consider the lowest excited odd bonds as following,\cite{lowest}
\begin{equation}
| \psi_f(2i-1,2i) \rangle =  \frac{1}{\sqrt{2}} | \uparrow_{2i-1} \uparrow_{2i} + \downarrow_{2i-1}  \downarrow_{2i} \rangle,
\end{equation}
and
\begin{equation}
| \psi_{af}(2i-1,2i) \rangle =  \frac{1}{\sqrt{2}} | \uparrow_{2i-1} \downarrow_{2i} + \downarrow_{2i-1}  \uparrow_{2i} \rangle,
\end{equation}
with bond energy $J_1+J_2$ and $-J_1+J_2$, respectively. Whereafter, we denote the one-particle excited state
\begin{equation}
| E(i) \rangle = | ... \phi_f \psi_f(2i-1,2i) \phi_f ... \rangle,
\end{equation}
for $J_1<0$; and
\begin{equation}
| E(i) \rangle = | ... \phi_{af} \psi_{af}(2i-1,2i) \phi_{af} ... \rangle,
\end{equation}
for $J_1>0$. That is, one odd bond $(2i-1, 2i)$ is in state $| \psi_{f(af)} \rangle$, while others remain in $| \phi_{f(af)} \rangle$. It is straightforward to verify that the transition matrix element of perturbation operator between $| E(i) \rangle$ and zeroth-order exact MPS ground state vanishes, i.e., $\langle \Psi_{\rm MPS} | L_1 \sigma_{2j}^z \sigma_{2j+1}^z | E(i) \rangle = 0$. This fact suggests that lowest one-particle excited states do not affect the MPS wavefunction in the first-order approximation, and only the multi-particle excited states or higher order perturbations will modify it. Further analysis shows that the perturbation term $L_1 \sigma_{2j}^z \sigma_{2j+1}^z$ will move the ``particle" along the chain, i.e., $\langle E(i) | L_1 \sigma_{2i}^z \sigma_{2i+1}^z | E(i+1) \rangle = L_1$, so the one-particle excitation dispersion up to the first-order approximation can determined as,\cite{Sachdev}
\begin{equation}
\epsilon_{k}^{\rm{I},\rm{II}} = 2J_2 + 2L_1\cos(k),
\label{eq-epsilon}
\end{equation}
in which $k= -\pi+\frac{2\pi}{N'}, ..., \pi-\frac{2\pi}{N'}, \pi$, and $| k \rangle = \frac{1}{\sqrt{N'}} \sum_{l=1}^{N'} e^{ikl} | E(l) \rangle$. This dispersion suggests that excitation gap of the system is nonzero in the phase diagram except for the line $J_2/L_1=1$ (gapless at $k=\pi$).

For regions III and IV, where the term $J_2 \sigma_{2i-1}^x \sigma_{2i}^x$ is regarded as perturbation, the same conclusion can be drawn after similar arguments, i.e., single-particle excited states will not modify the MPS ground state up to the first-order single-particle perturbation, and MPS is also a very nice approximation in these two gapped regions. Which is different, in this case the excited particle is revealed to be a moving ``domain wall'' instead of a single excited odd bond, and the dispersion relation can be verified as $\epsilon_{k}^{\rm{III}, \rm{IV}} = 2L_1 + 2J_2\cos(k)$.

\subsection{iTEBD and imaginary time projections}

Beyond the perturbation arguments, imaginary time projection technique iTEBD is employed to accurately determine the variational MPS wavefunction.\cite{Vidal}
To be concrete, the variational ground state $| \Psi_g \rangle$ (in MPS form) can be obtained by acting the imaginary time evolution operator exp(-$\beta \hat{H}$) on an arbitrary initial state $| \Psi_0\rangle$. The operator $\exp(-\beta \hat{H})$ is expanded through Suzuki-Trotter decomposition as a sequence of two-site gates $U^{[i,i+1]} = \exp(-\tau \hat{h}_{i,i+1})$, where $\hat{h}_{i,i+1}$ is the local bond Hamiltonian, and $\tau$ means small Trotter step length. In the limit $\beta \rightarrow \infty$, the resulting wave function exp(-$\beta \hat{H}$)$| \Psi_0\rangle$ will converge (or be very close) to the ground state of $\hat{H}$. Fig. \ref{fig02} illustrates calculation errors compared with the exact solutions. Some typical parameters including critical and noncritical points are concerned. The errors converge rapidly with enhancing $\chi$, in noncritical regions very accurate results can be obtained even with the smallest nontrivial bond dimension $\chi=2,4$, which convince us that MPS description of the present system is not only adequate but also highly efficient and accurate. In practical implementations, the convergence of results with different bond dimension $\chi$ has always been checked, and for most cases up to $\chi=40$ is quite enough. The total number of iterations taken is about $10^{5} \sim 10^{6}$. We first start with a step $\tau=10^{-1}$, and then diminish it to $\tau=10^{-8}$ gradually. Whenever $\tau$ is small enough, this procedure would bring the MPS to its canonical form, which would be useful for calculating the entanglement entropies, as well as the local observables including energy per site and local magnetizations, \emph{etc}.

During the iTEBD process for two-period MPS (Eq. (\ref{MPS})), only four tensors ($T_a$, $T_b$, $\Lambda_a$, and $\Lambda_b$) are involved and updated in each iteration step. In order to capture more symmetry broken phases with larger unit cell, sometimes we need four-period MPS which includes eight different tensors ($T_a$, $T_b$, $T_c$, $T_d$, $\Lambda_a$, $\Lambda_b$, $\Lambda_c$, and $\Lambda_d$). For example, region III in Fig. \ref{Fig1} is verified as a stripe AF ordered phase (one such ordered spin configuration is illustrated in Fig. \ref{fig01} (a)), and this stripe AF order can be well described with four-period MPSs, but not two-period ones.

\subsection{Quantum Entanglement and Fidelity}
Quantum entanglement has close relationship with QPTs in many-body
systems,\cite{Vidal-QPT} and much effort has been devoted to
studying the quantitative description of entanglement in quantum
systems.\cite{Schlientz,Bennett,Bennett2,Osterloh,Verstraete,Amico,Huang}
In order to describe the QPTs in the EQCM, the von Neumann
entropy $S_{\rm{vN}}$ is adopted as a bipartite entanglement measure.\cite{Amico}
When the MPS is gauged to its canonical form, i.e., we can cut an
arbitrary bond in the system, and obtain a Schmidt decomposition
as,
\begin{equation}
| \Psi \rangle = \sum_{\alpha=1}^{\chi} | \Phi^L_{\alpha} \rangle \Lambda_{\alpha} | \Phi^R_{\alpha} \rangle.
\label{eq-canonical}
\end{equation}
Here, $| \Phi^L_{\alpha} \rangle$ ($| \Phi^R_{\alpha} \rangle$) represent the orthonormal bases of subsystem to the left (right) of the broken bond, and $\Lambda$ is a diagonal matrix.
Correspondingly, the canonical MPS would satisfy the following two equations,
\begin{eqnarray}
\sum_m \sum_{\alpha} (\Gamma^{\ast})^{m}_{\alpha, \beta'} \Lambda^2_{\alpha} \Gamma^m_{\alpha, \beta''} & = & \delta_{\beta'  \beta''}, \nonumber \\
\sum_m \sum_{\beta} (\Gamma^{\ast})^{m}_{\alpha', \beta} \Lambda^2_{\beta} \Gamma^m_{\alpha'', \beta} & = & \delta_{\alpha'  \alpha''}.
\label{eq-canonical-condition}
\end{eqnarray}
The superscript $^{\ast}$ means complex conjugate. It is easy to check that in the above limit
cases $L_1=0$, the exact MPSs satisfy Eq. (\ref{eq-canonical-condition}),
and they are thus in canonical form. Given the MPS in its canonical form,
bipartite entanglement of half chain ($S_{\rm{half}}$) can be directly read
from diagonal matrix $\Lambda$ (see Eq. (\ref{eq-canonical})),
\begin{eqnarray}
S_{\rm{half}} & = & - \rm{Tr}(\Lambda^2 \rm{log_2} \Lambda^2) \notag \\
              & = & - \sum_{\alpha=1}^{\chi} \Lambda_{\alpha}^{2} {\rm log_2} \Lambda_{\alpha}^{2}.
\label{Definition Of Entanglement}
\end{eqnarray}
Notice, for two-period MPS, we can define two different bipartite entanglement
measures $S_{2i-1,2i} = - \rm{Tr} (\Lambda_{a}^{2} {\rm log_2} \Lambda_{a}^{2})$ and $S_{2i,2i+1} = - \rm{Tr} (\Lambda_{b}^{2} {\rm log_2} \Lambda_{b}^{2})$, on odd and even bonds, respectively.
Besides $S_{\rm{half}}$, people are also interested in the block entanglement,
which is defined as follows,
\begin{equation}
S_{\rm L} = - \textrm{Tr} [\rho_{\rm L} {\rm log_2}(\rho_{\rm L}) ]=
- \textrm{Tr} [\rho_{\rm env_L} {\rm log_2}(\rho_{\rm env_L})],
\end{equation}
where $\rho_{\rm L}$ is the reduced density matrix of the $L$ spin system, and $\rm{env_L}$ means the environment (rest of the chain). $S_{\rm L}$ characterizes the entanglement between $L$ adjacent spins and the environment. In practical calculations, the density matrix $\rho_{\rm{env_L}}$ is supported by $\chi$ Schmidt bases, and employed to calculate the block entanglement entropy for arbitrary spin portion $L$. We would like to stress that the two kinds of entanglement measures are both
von Neumann entropies just that the bipartition happens to be between the left and right halves in the first case and between a block and the rest in the second case.

Except for the entanglement measures, fidelity per site $f$ is also used to detect the QPTs,\cite{Zhou} which is defined as
\begin{equation}
f = \lim_{N\rightarrow \infty} \frac{\langle \Psi | \Psi_{\rm ref}
\rangle}{N}. \label{eq-fidelity per site}
\end{equation}
$| \Psi \rangle$ is the ground-state wavefunction of the present system, and $| \Psi_{\rm ref} \rangle$ is a reference state, $f$ indicates how fast the overlap of two distinct state decays to zero with increasing the length of the chain. The bifurcation and singular points of $f$ can be utilized to locate the QPTs.\cite{Zhou, Wanghl}

Remarkably, the von Neumann entropy and the fidelity per site $f$ can be
conveniently obtained in the framework of MPS.
Therefore, we will adopt them, along with the energy, magnetization,
and nearest-neighbor correlators, to study the phase transitions of 1D EQCM in the following sections.

\section*{III. QPTs in the one-dimensional EQCM}

\subsection{Ground State Energy, Entanglement Entropy, and Local Order Parameter}
\label{sec-OrderPara}
\begin{figure}
\begin{center}
\includegraphics[width=1.0\linewidth]{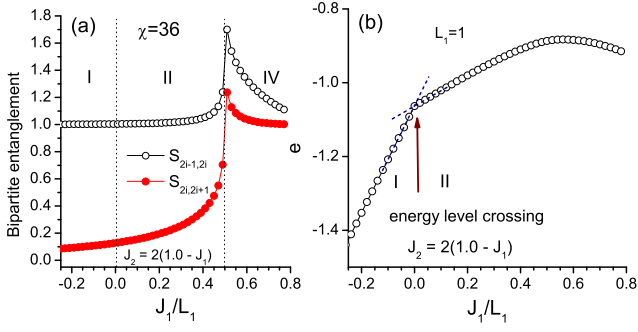}
\end{center}
\caption{(Color online) (a) Half-chain entanglement entropy on odd bond $S_{2i-1, 2i}$ and even bond $S_{2i, 2i+1}$. (b) Ground state energy per site, two dotted lines represent energies of the adiabatically evolved states from left and right sides of the first-order QPT point.}
\label{Fig2}
\end{figure}

Firstly, we consider the QPTs along the line $J_2=2.0\times(1.0-J_1)$
(dash line (a) in Fig. \ref{Fig1}). As the phase diagram illustrates, with increasing $J_1$,
the system should undergo two sequential QPTs: one first-order QPT from region I to region II and then the second-order one from region II to region IV. The bipartite entanglement entropies $S_{2i-1,2i}$ and $S_{2i, 2i+1}$ are plotted in Fig. \ref{Fig2} (a). From Fig. \ref{Fig2} (a), it is clearly seen that there exists only one singular point $J_1 = 0.5$ (and $J_2 = 1.0$) where a second-order QPT takes place. From Fig. \ref{Fig2} (b), an energy level crossing happens at $J_1 = 0$, which indicates that a first-order QPT should occur there. However, as shown in Fig. \ref{Fig2} (a), the bipartite entanglement changes continuously across the first-order QPT.\cite{explan-Ent} Therefore, the first-order QPT in EQCM is missed by the entanglement measurement $S_{\rm{half}}$. Notice that the adiabatic continuations are plotted with dashed lines in Fig. \ref{Fig2} (b), which illustrate the adiabatically evolved states from the left (or right) of the transition point, explicitly revealing the nature of level crossing at the first-order QPT. \cite{orus-2dqcm}

\begin{figure}
\begin{center}
\includegraphics[width=1.0\linewidth]{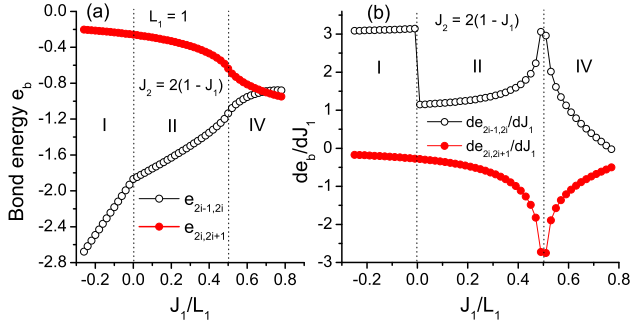}
\end{center}
\caption{(Color online) (a) The odd (even) bond energy $e_b$ and (b) their first-order derivatives along the line $J_2 = 2(1-J_1)$.}
\label{Fig3}
\end{figure}

Next, we pay attention to the ground-state energy on odd and even
bonds (denoted as $e_{2i-1,2i}$ and $e_{2i,2i+1}$, respectively) and
their first-order derivatives (see Fig. \ref{Fig3} (a) and (b)). We
find that, the first-order QPT at $J_1 = 0$ can be detected by the
energy level crossing of the odd bond energy (Fig. \ref{Fig3} (a))
or the discontinuous behavior of its first-order derivative (Fig.
\ref{Fig3} (b)). Furthermore, the singular behavior of the
first-order derivatives (of both $e_{2i-1,2i}$ and $e_{2i,2i+1}$) at
$J_1 = 0.5$ indicates the occurrence of the second-order QPT.
According to the Feynman-Hellmann theorem
\begin{eqnarray}
\frac{\partial
e}{\partial \lambda} = \langle \psi|\frac{\partial
\hat{H(\lambda)}}{\partial \lambda}|\psi\rangle,
\end{eqnarray}
where $\lambda$ is a
tunable parameter in the Hamiltonian. One can speculate that the
first-order derivative of bond energy is in fact a second-order
derivative of site energy $e$. Take even bond
energy $e_{2i,2i+1}$ as an example, $ de_{2i,2i+1}/dJ_1 =
 d^2 e/(dJ_1dL_1)$, and it is thus expected to show singular
behaviors around the second-order QPTs (as Fig. \ref{Fig3} (b)
shows).

\begin{figure}
\begin{center}
\includegraphics[width=0.95\linewidth]{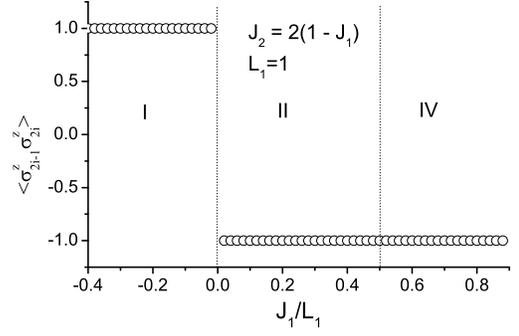}
\end{center}
\caption{Nearest neighbor correlator $\langle \sigma^{z}_{2i-1}\sigma^{z}_{2i}\rangle$.}
\label{Fig4}
\end{figure}

\begin{figure}
\begin{center}
\includegraphics[width=1.0\linewidth]{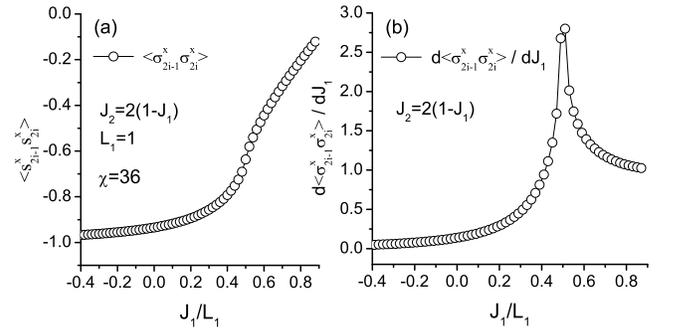}
\end{center}
\caption{Short-range correlation $ \langle \sigma^{x}_{2i-1}\sigma^{x}_{2i}\rangle$ (a), and its first-order derivative (b).}
\label{Fig5}
\end{figure}

On the other hand, $ de_{2i-1,2i}/dJ_1$  =
$\langle\sigma_{2i-1}^z \sigma_{2i}^z\rangle$ + $J_2 d \langle
\sigma_{2i-1}^x \sigma_{2i}^x \rangle/dJ_1 $, and the short-range
correlators $\langle \sigma^{z}_{2i-1}\sigma^{z}_{2i}\rangle$ and
$\langle \sigma^{x}_{2i-1}\sigma^{x}_{2i}\rangle$ on odd bonds are
calculated and shown in Figs. \ref{Fig4}, \ref{Fig5}. From Fig.
\ref{Fig4}, we find that the short-range correlation
$\langle\sigma^{z}_{2i-1}\sigma^{z}_{2i}\rangle$ is $+1$ in region
I, but abruptly changes into $-1$ as entering into regions II and
IV. So, the first-order QPT takes place with a sign change of
two-site correlation function $\langle\sigma^{z}_{2i-1}\sigma^{z}_{2i}\rangle$ on odd bonds, and
causes a discontinuity in $de_{2i-1,2i}/dJ_1$ curve. In Fig. \ref{Fig5} (a), the two-site correlator
$\langle\sigma^{x}_{2i-1}\sigma^{x}_{2i}\rangle$ on odd bonds behaves continuously, 
and the divergent peak of its first-order derivative (Fig. \ref{Fig5} (b)) can signal the critical point ($J_{1} = 0.5$).
Consequently, derivative $de_{2i-1,2i}/dJ_1$ also has a divergent peak at $J_{1} = 0.5$. At last, it is worth noticing that, although derivatives
$de_{2i-1,2i}/dJ_1$ and $de_{2i,2i+1}/dJ_1$ are both divergent at
$J_{1} = 0.5$, they have different signs and cancel with each other,
the first-order derivative $de/dJ_1 = d(e_{2i-1,2i} +
e_{2i,2i+1})/dJ_1$ is continuous across the second-order QPT point.

\begin{figure}
\begin{center}
\includegraphics[width=0.9\linewidth]{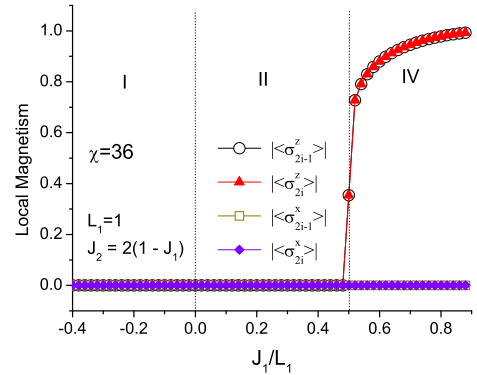}
\end{center}
\caption{(Color online) Local magnetizations $| \langle \sigma^{x}\rangle |$ and $| \langle \sigma^{z}\rangle |$.}
\label{Fig6}
\end{figure}

To attain comprehensive understandings of the QPTs in EQCM, we also
compute the local magnetizations $| \langle \sigma^{x}\rangle |$ and
$| \langle \sigma^{z} \rangle |$ (shown in Fig. \ref{Fig6}), whose
values are independent of the sites owing to the translational invariant MPS.
It is observed that across the critical point $J_{1} = 0.5$, the EQCM goes into a region (IV) with nonzero
$|\langle \sigma^{z} \rangle |$, and their values have different
signs on odd and even sites, i.e., staggered magnetization, and thus
region IV can be recognized as a semi-classical N\'{e}el phase.
Thus, the magnetization $|\langle \sigma^{z}\rangle|$ (to be more
strict, staggered magnetization $M^z_{\rm{Neel}} = \frac{1}{2} |\langle
\sigma_{2i-1}^z - \sigma_{2i}^z \rangle |$) can be recognized as the
local order parameter characterizing region IV in Fig. \ref{Fig1},
however, the regions I and II are both disordered phases, and can
not be distinguished by the local order parameters.\cite{explan-OP}
On the other hand, Fig. \ref{Fig6} reveals that $| \langle
\sigma^{x} \rangle |$ vanishes in regions I, II, and IV on either odd
or even sites.

\begin{figure}
\begin{center}
\includegraphics[width=1.0\linewidth]{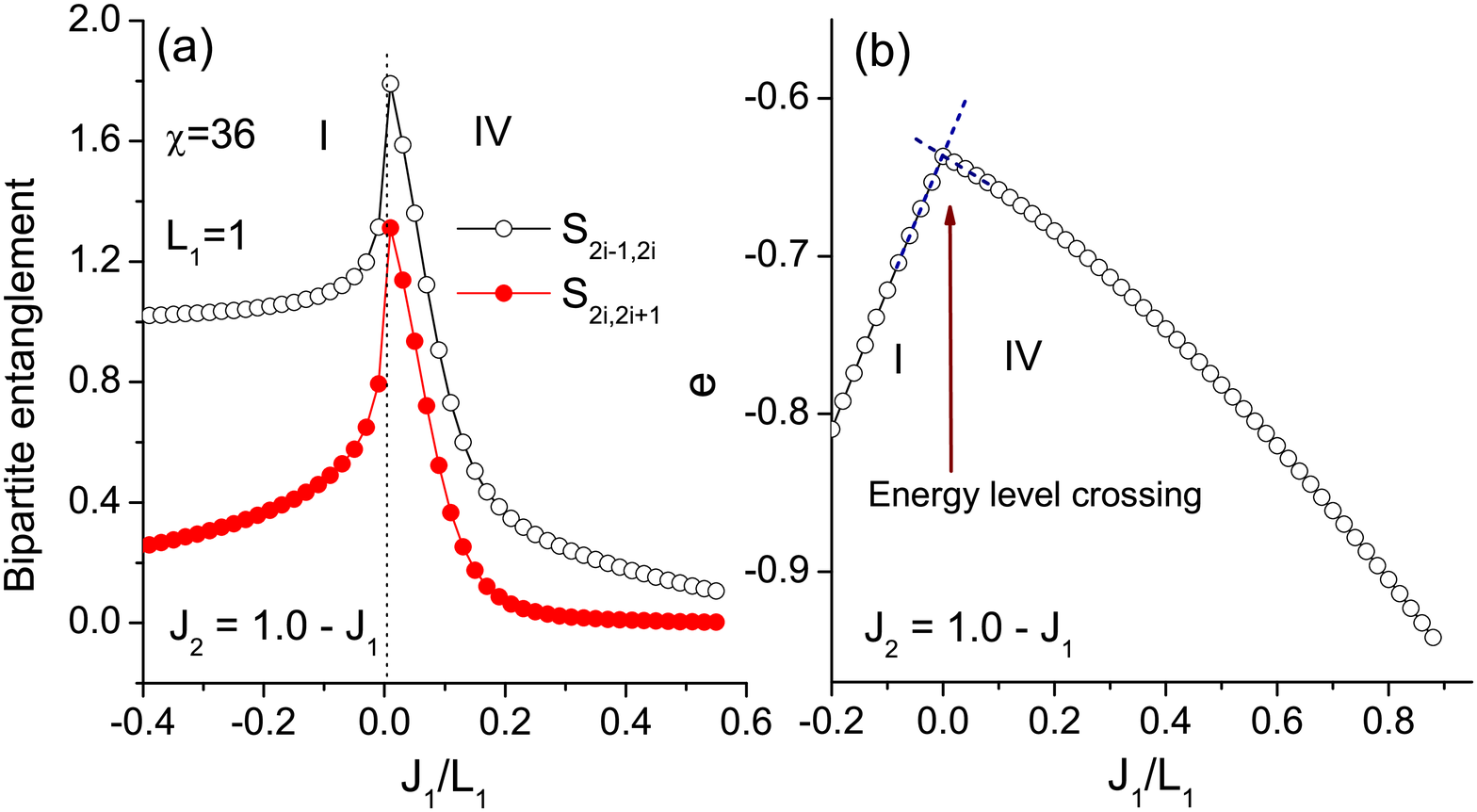}
\end{center}
\caption{(Color online) (a) Bipartite entanglement measures $S_{2i-1,2i}$ and $S_{2i,2i+1}$ along the line $J_2 =1.0 - J_1$. (b) Ground state energy per site $e$, the dotted lines represent energy of adiabatically continued states.}
\label{Fig7}
\end{figure}

In order to discuss the bipartite entanglement behavior across the multicritical point ($J_1=0, J_2=1.0$), we consider the line $J_2 = 1.0 - J_1$. Along this line, with increasing $J_1$, the ground state of EQCM will go from region I into region IV through the multicritical point. The odd and even bond bipartite entanglement measures are plotted in Fig. \ref{Fig7} (a). It is noticed that the QPT can be recognized by the sharp peaks of $S_{2i-1,2i}$ and $S_{2i,2i+1}$, which confirms it as a quantum critical point. It should also be mentioned that, except for the similar second-order QPT character in entanglement measure, a distinctive ground-state energy level crossing is also clearly shown in Fig. \ref{Fig7} (b), where adiabatic continuations are again employed to verify this conclusion. Therefore, both the first- and second-order QPT features at this multicritical point are revealed by our calculations. Besides, the local magnetization $| \langle \sigma^{x} \rangle |$ and $| \langle \sigma^{z} \rangle |$ in regions I and IV are also evaluated (not shown for the sake of space), and similar behaviors from disordered region I to N\'{e}el ordered region IV as discussed above are again observed.

\begin{figure}
\begin{center}
\includegraphics[width=1.0\linewidth]{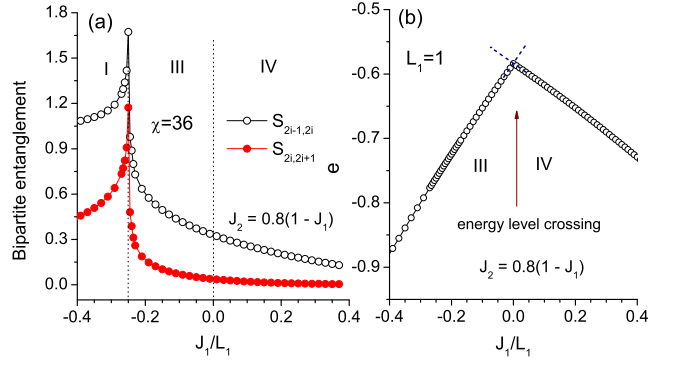}
\end{center}
\caption{(Color online) (a) Bipartite entanglement on odd bond $S_{2i-1,2i}$ and even bond $S_{2i,2i+1}$ along line $J_2 =0.8(1.0 - J_1)$.\cite{explan-four-period} (b) Ground state energy per site $e$ (the dotted lines are adiabatic continuations).}
\label{Fig8}
\end{figure}

\begin{figure}
\begin{center}
\includegraphics[width=1.0\linewidth]{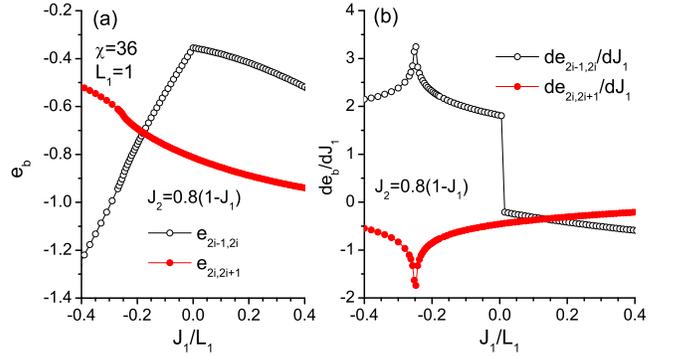}
\end{center}
\caption{(Color online) (a) Ground state energy of odd and even bonds and (b) their first derivatives.}
\label{Fig9}
\end{figure}

\begin{figure}
\begin{center}
\includegraphics[width=1.0\linewidth]{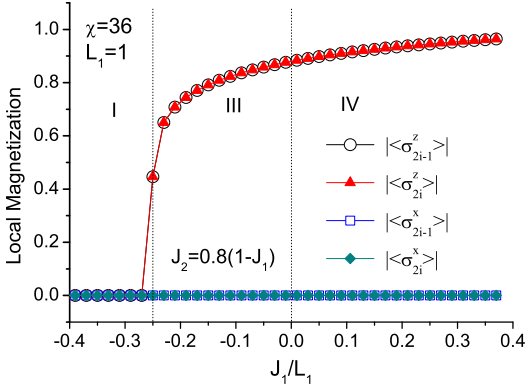}
\end{center}
\caption{(Color online) Local magnetization $| \langle \sigma^{x}\rangle |$ and $| \langle \sigma^{z}\rangle |$ ($J_2 = 0.8(1-J_1)$). The magnetic order is stripe AF in region III, and N\'{e}el AF in region IV.}
\label{Fig10}
\end{figure}

Lastly, we consider the QPTs along the line $J_2=0.8 \times
(1.0-J_1)$.  With increasing $J_1$, the ground state of EQCM will go
from region I to region III, and then enter into region IV. The
bipartite entanglement on odd and even bonds,
and the ground-state energy per site $e$ are plotted in Fig. \ref{Fig8} (a) and (b), respectively.
We find that, although the second-order QPT at $J_1 =-0.25, J_2=1$
is signaled by a singular peak of the entanglement entropy, the
first-order QPT at $J_1 =0$ with distinct ground-state energy level
crossing (Fig. \ref{Fig8} (b)) is again missed by the entanglement
measure (Fig. \ref{Fig8} (a)). However, as shown in Fig. \ref{Fig9}
(a) and (b), the bond energy and their first-order derivatives are
able to capture all the QPTs. In addition, magnetization is
calculated and shown in Fig. \ref{Fig10}, nonzero $| \langle
\sigma^{z}\rangle |$ is found in region III and IV, and $| \langle
\sigma^{x} \rangle |$ vanishes along the whole line on either odd or
even sites.

Moreover, in Fig. \ref{Fig10}, although the magnitude of $| \langle \sigma^z \rangle |$ change smoothly through the phase transition point $J_1=0, J_2=0.8$, the magnetic order is quite different in region III with that of region IV. Calculations indicate that, the correlators $\langle\sigma^{z}_{2i-1}\sigma^{z}_{2i}\rangle = 1$ in region III, show distinct difference with those in N\'{e}el phase (region IV), where $\langle\sigma^{z}_{2i-1}\sigma^{z}_{2i}\rangle = -1 $. In fact, the magnetic order in region III is four-period stripe AF order, quite different from N\'{e}el order in region IV. In N\'{e}el phase, the spins are arrayed in ``up-down-up-down" pattern (see Fig. \ref{fig01} (b)); while in stripe AF phase, they are in `` up-up-down-down '' arrangements, one typical spin configuration of such phase is illustrated in Fig. \ref{fig01} (a). The stripe AF order in 1D EQCM was previously proposed in Ref. \onlinecite{Mahdavifar} with finite-size calculations by Lanczos method, and it is confirmed here by our results directly in the thermodynamic limit.

As the N\'{e}el order parameter $M_{\rm{Neel}}^z$ defined in region IV,
$M_{\rm{stripe}}^z = \frac{1}{2} | \langle \sigma^z_{4n-3} -
\sigma^z_{4n-1} \rangle |$, where $n=1,2,...,N'/2$, can be defined
as the local order parameter in region III. Then, $M_{\rm{Neel}}^z$ is
nonzero in region IV, and vanishes abruptly in region III; while the
reverse is $M_{\rm{stripe}}^z$, which appears in region III, and drops to
zero in region IV. Therefore, the first-order QPT between regions
III and IV can be recognized by evaluating local order parameters,
quite different with the transition between regions I and II
discussed above.

\subsection{Block Entanglement Entropy}

\begin{figure}
\begin{center}
\includegraphics[width=1.0\linewidth]{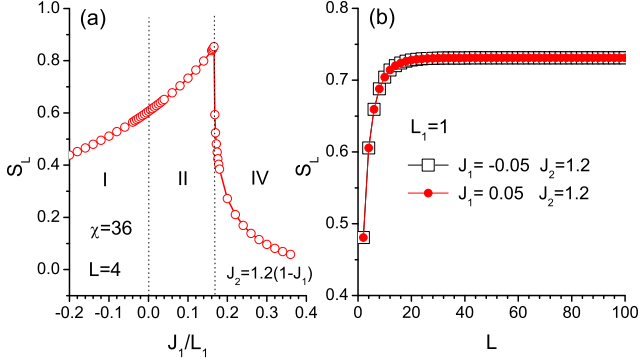}
\end{center}
\caption{(Color online) (a) Entanglement entropy $S_{\rm L}$ (with block size $L=4$) along line $J_{2}=1.2(1.0-J_{1})$ and (b) $S_{\rm L}$ saturates rapidly with increasing $L$ in the vicinity of the first-order QPT line.}
\label{Fig11}
\end{figure}

Besides the half chain entanglement, the block entanglement entropy $S_{\rm L}$ are also calculated, which provides a measurement of the amount of entanglement between $L$ adjacent spins and the rest of the system (environment). With MPS wavefunction, we are able to obtain the reduced density matrix of the environment supported by the bond bases, and hence can calculate the $S_{\rm L}$ with length $L$ up to several hundreds of sites at ease. The block entanglement entropy ($S_{\rm L}$)
with $L=4$ along the line $J_{2}=1.2 \times (1.0-J_{1})$ is plotted in Fig. \ref{Fig11} (a). With increasing $J_1$, two sequential QPTs will take place: one first-order QPT from region I to region II and the other second-order QPT from region II to region IV. However, from Fig. \ref{Fig11} (a), we find that only the second-order QPT at $J_2 = 1$ can be detected by the peak of the block entanglement entropy $S_{\rm L}$, the first-order QPT (at $J_1 = 0$, $J_2 = 1.2$) between phase I and phase II is missed again. $S_{\rm L}$ continuously approaches the same value whether $J_1 \rightarrow 0^-$ or $J_1 \rightarrow 0^+$ with fixed $J_2$. In Fig. \ref{Fig11} (b), for the non-critical ground state, when block size $L$ increases, the block entropy $S_{\rm L}$ enhances and quickly becomes saturated, well satisfying the entanglement area law.\cite{Eisert} These observations on the block entanglement are consistent with those proposed in Ref. \onlinecite{Eriksson}. Therefore, the entanglement measures, including block entanglement entropy $S_{\rm L}$ and half chain entanglement $S_{\rm{half}}$, are indeed not able to detect the first-order QPTs in 1D EQCM.

\begin{figure}
\begin{center}
\includegraphics[width=1.0\linewidth]{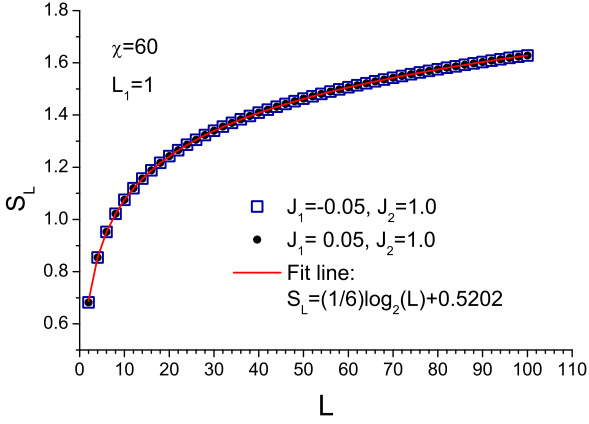}
\end{center}
\caption{(Color online) Scaling of the block entanglement entropy $S_{\rm L}$ on the second-order QPT line $J_{2}=1.0$, the solid line is fit to numerical data.}
\label{Fig12}
\end{figure}

Next, the scaling behavior of the block entropy $S_{\rm L}$ on the second-order QPT line $J_2 = 1$ is investigated. As shown in Fig. \ref{Fig12}, the block entropy $S_{\rm L}$ exhibits divergent behavior with increasing block size. As derived in Ref. \onlinecite{Holzhey}, in a $1+1$ dimensional conformal field theory, the entropy of a subregion of length $L$ reads
\begin{equation}
S_{\rm L}=\frac{c+\bar{c}}{6}{\rm log}_{2}(L)+ k,
\label{SL}
\end{equation}
with a coefficient given by the holomorphic and antiholomorphic
central charges $c$ and $\bar{c}$ of the theory. From Fig. \ref{Fig12}, we find that the divergent $S_{\rm L}$ on the second-order QPT line can be well fitted by $S_{\rm L}=\frac{1}{6}{\rm log}_{2}(L)+0.5202$, with central charges c = $\bar{c}$=1/2, i.e., the $S_{\rm L}$ displays a logarithmic divergence on the second-order QPT line. Therefore, we disclose that the critical behavior of EQCM can be described by a free fermionic field theory,\cite{Vidal-QPT} with central charges $c_f$ = $\bar{c}_f$=$1/2$.

\subsection{Fidelity Calculations}

Except for the entanglement, the fidelity measure defined in Eq. (\ref{eq-fidelity per site}) is also utilized to study the QPTs in EQCM. Facilitated with MPS framework, it is straightforward that $f$ can be obtained by evaluating the maximum eigenvalue of the transfer-matrix defined as
\begin{equation}
P_{\alpha'\alpha,\beta'\beta} = \sum_{m} \widetilde{\Lambda}_{\alpha'} (\widetilde{\Gamma}^{\ast})_{\alpha',\beta'}^{m} \Lambda_{\alpha} \Gamma^m_{\alpha, \beta},
\end{equation}
in which $\widetilde{\Gamma}$ (along with $\widetilde{\Lambda}$) represents the reference state. Considering the multi-period MPS wavefunctions (period 2 for regions I, II, and IV, and period 4 for region III), we slightly modify it and define the fidelity per unit cell, which is the maximum eigenvalue of the transfer-matrix defined in a unit cell. For instance, the transfer matrix of two-period MPS can be defined as following,
\begin{eqnarray}
P^{a,b}_{\alpha'\alpha, \gamma'\gamma} & = & \sum_{\beta',\beta,m_{2i-1},m_{2i}}  (\widetilde{\Lambda}_a)_{\alpha'} (\widetilde{T}_a^{\ast})^{m_{2i-1}}_{\alpha',\beta'} (\widetilde{\Lambda}_b)_{\beta'} (\widetilde{T}_b^{\ast})^{m_{2i}}_{\beta',\gamma'} \nonumber \\
&\times&(\Lambda_a)_{\alpha} (T_a)^{m_{2i-1}}_{\alpha,\beta} (\Lambda_b)_{\beta} (T_b)^{m_{2i}}_{\beta,\gamma},
\end{eqnarray}
which is a $\chi_a^2 \times \chi_a^2$ matrix. The transfer matrix of four-period MPS can be similarly written down.

\begin{figure}
\begin{center}
\includegraphics[width=1.0\linewidth]{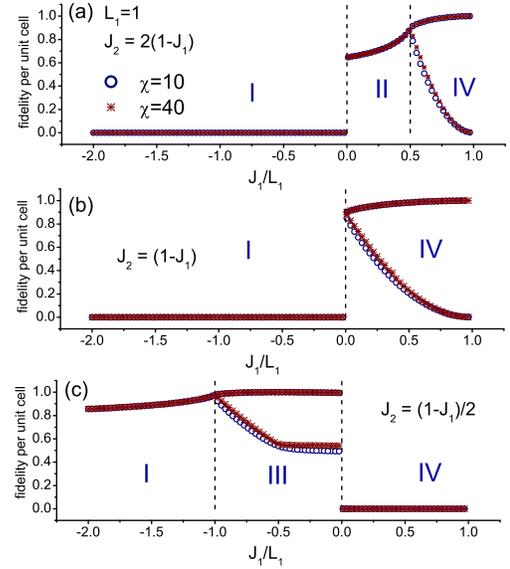}
\end{center}
\caption{(Color online) Fidelity per unit cell along three different lines are present in (a), (b), and (c). The discontinuities of the curves indicate the occurrence of first-order QPTs, and the bifurcation phenomena manifest the spontaneous $Z_2$ symmetry breaking, where second-order QPTs take place. Four-period MPS is adopted during the calculations. Notice that the reference state in (c) is different from that in (a) and (b), see the text for more information.}
\label{Fig13}
\end{figure}

In Fig. \ref{Fig13}, the results of fidelity per unit cell are present (the MPSs are generally set as period 4) along three different lines, $J_2=2(1-J_1)$, $J_2=1-J_1$, and $J_2=(1-J_1)/2$, respectively. In Fig. \ref{Fig13} (a), the line traverses regions I, II, and IV, and in Fig. \ref{Fig13} (b), regions I and IV are involved. During these calculations, the ground state of Hamiltonian Eq. (\ref{Hamiltonian}) with parameter $J_1=1, J_2=0$ is set as the reference state (i.e., an Ising AF state). Owing to the spontaneous $Z_2$ symmetry breaking in the N\'{e}el phase, $f$ shows bifurcation behaviors in Fig. \ref{Fig13} (a) and (b), and the bifurcation points locate the second-order QPTs. Besides, the first-order QPTs can also be recognized from the discontinuities in fidelity curves. It is worth noticing that the results in Fig. \ref{Fig13} (b) again reveal multicritical properties of the transition occurred at $J_1=0, J_2=1$, i.e., the discontinuity of $f$ indicates a first-order QPT, while the bifurcation phenomenon reveals second-order QPT character. In Fig. \ref{Fig13} (c), we choose $J_1=-0.5, J_2=0.75$ as the reference point, the bifurcation at $J_1=-1$ indicates second-order QPT between regions I and III, and the discontinuity at $J_1=0$ suggests the first-order QPT between stripe AF and N\'{e}el phases (regions III and IV, respectively).

\subsection{Entanglement Spectrum, Dual Transformation, and String Order Parameters}
\label{SOP}

Through previous analysis in subsection \ref{sec-OrderPara}, it is uncovered that no local order parameter can be utilized to distinguish the two disordered phases regions I and II in Fig. \ref{Fig1}, as well as to detect the first-order QPTs between them. In this subsection, the non-local string order parameters in regions I and II are computed and discussed.

\begin{figure}
\begin{center}
\includegraphics[width=1.0\linewidth]{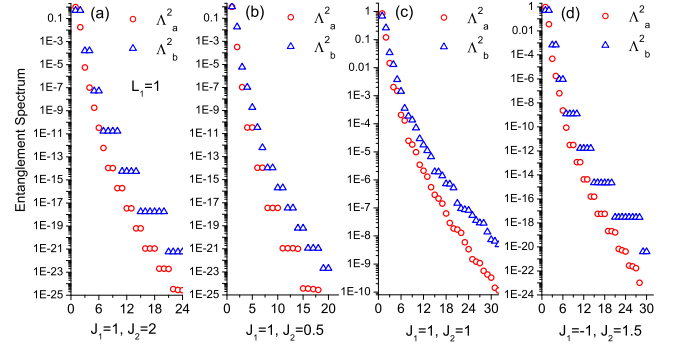}
\end{center}
\caption{(Color online) The entanglement spectra of several noncritical (a, b, and d) and critical (c) points. In (a) and (d), $\Lambda_b^2$'s are doubly degenerate.}
\label{Fig14}
\end{figure}

In Fig. \ref{Fig14}, several typical entanglement spectra of 1D EQCM
are shown, which exhibit the eigenvalues of the reduced density
matrix of half-infinite chain by dividing the system via any bond.
For the canonical MPS, entanglement spectrum can be recognized as the diagonal elements of $\Lambda^2$ in Eq. (\ref{eq-canonical}).
For the present two-period system, it is free to cut an even or odd bond, thus we have two entanglement
spectrums ($\Lambda_a^2$ and $\Lambda_b^2$) for a single parameter
point. In Fig. \ref{Fig14} (a), (b), and (d), non-critical points
are concerned, and the eigenvalues decay roughly exponentially;
while in Fig. \ref{Fig14} (c), for critical point, the entanglement
spectrum decays much slower, and in some algebraical way. Another
distinct feature is the doubly degenerate $\Lambda_b^2$ for
disordered phases I and II (Fig. \ref{Fig14} (a) and (d)), which
implies the existence of the non-local string order
parameters.\cite{Pollmann}

Previous studies suggested that along the line $J_1=0$ there exist
two topological distinct disordered phases for $J_2/L_1>1$ and
$J_2/L_1<1$, and the phase transition between them (at $J_2/L_1=1$)
is disclosed as a topological QPT,\cite{Feng} characterized by
non-local string order parameters. Other than this disordered line,
it is an interesting question that whether the string order
parameters in regions I and II still exist or not. To accomplish
this task, standard Kramers-Wannier dual
transformation\cite{Kramers} is employed to map the present model to
the quantum-Ising system. The dual mapping of each terms in
Hamiltonian Eq. (\ref{Hamiltonian}) are as followings (here we adopt
the formalism introduced in Ref. \onlinecite{Kohmoto}, and a
permutation of even and odd bonds is taken before dual
transformation),
\begin{eqnarray}
J_1 \sigma_{2i}^z \sigma_{2i+1}^z & \rightarrow & -J_1 \tau_{i}^z \tau_{i+1}^z, \nonumber \\
J_2 \sigma_{2i}^x \sigma_{2i+1}^x & \rightarrow & -J_2 \widetilde{\sigma}_{i}^z \widetilde{\sigma}_{i+1}^z, \nonumber \\
L_1 \sigma_{2i-1}^z \sigma_{2i}^z & \rightarrow & L_1 \widetilde{\sigma}_{i}^x,
\end{eqnarray}
and thus the Hamiltonian is as
\begin{equation}
\widetilde{H} = \sum_{i=1}^{N'} -J_1 \tau_{i}^z \tau_{i+1}^z - J_2 \widetilde{\sigma}_{i}^z \widetilde{\sigma}_{i+1}^z + L_1 \widetilde{\sigma}_i^x,
\label{eq-dual-Hamiltonian}
\end{equation}
where $\widetilde{\sigma}$ and $\tau$ are Pauli matrices on the dual
lattice. Dual Hamiltonian Eq. (\ref{eq-dual-Hamiltonian}) can be regarded as two decoupled Ising spin chains (couplings $-J_1$ and $-J_2$, respectively),\cite{Perk}
and the $\widetilde{\sigma}$ chain is under transverse field ($L_1 \widetilde{\sigma}_i^x$ term). There may exist two types of spontaneous long-range orders, i.e., $\langle
\widetilde{\sigma}_k^z \widetilde{\sigma}_n^z \rangle$ and $\langle
\tau_k^z \tau_n^z \rangle$, which can be mapped back to the original
system as the following non-local string order parameters,
\begin{eqnarray}
(-1)^{n-k} \langle \tau_k^z \tau_n^z \rangle & \rightarrow & \langle \sigma_{2k}^z \sigma_{2k+1}^z ... \sigma_{2n-2}^z \sigma_{2n-1}^z \rangle,  \nonumber \\
(-1)^{n-k} \langle \widetilde{\sigma}_k^z \widetilde{\sigma}_n^z \rangle & \rightarrow & \langle \sigma_{2k}^x \sigma_{2k+1}^x ... \sigma_{2n-2}^x \sigma_{2n-1}^x \rangle.
\end{eqnarray}
The two types of $\sigma$ operator strings can be denoted as $O^{zz}(n-k)$ and $O^{xx}(n-k)$, respectively. Owing to the absence of transverse field on $\tau$ spins in the dual model Eq. (\ref{eq-dual-Hamiltonian}), $O^{zz}(n-k)$ is always nonzero in the whole phase diagram. To be specific, it is found that $O^{zz}(n-k) = 1$ for region I (and also region III), while $O^{zz}(n-k) = (-1)^{n-k}$ for region II (and IV). This is owing to that in the dual model there exists ferromagnetic long range order ($\langle \tau^z_{k} \tau^z_{n} \rangle = 1$) for $J_1>0$, and AF long range order ($\langle \tau^z_{k} \tau^z_{n} \rangle = (-1)^{n-k}$) for $J_1<0$. On the other hand, this conclusion can be easily verified by noticing the nearest neighbor correlators $\langle \sigma_{2i-1}^z \sigma_{2i}^z \rangle = 1$ for region I (III) and $-1$ for region II (IV), which can also be regarded as good quantum numbers for ground states.

\begin{figure}
\begin{center}
\includegraphics[width=1.0\linewidth]{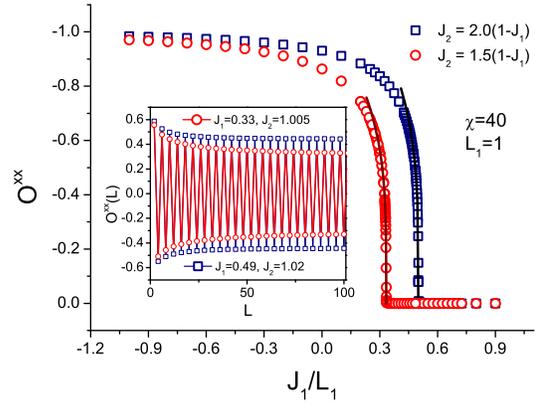}
\end{center}
\caption{(Color online) The nonlocal order parameter $O^{xx}$, inset
shows the behaviors of $O^{xx}(L)$ with portion length $L$ (shown
for every two sites). The converged value $O^{xx}(\infty)$ is
nonzero in disordered regions I and II, and vanishes in ordered
phases III and IV. The fitting lines illustrate the asymptotic
behavior $|O^{xx}(\infty)| \sim (1-J_2^2)^{1/4}$ in the vicinity of
the second-order QPT.} \label{Fig15}
\end{figure}

The behavior of the other string order parameter $O^{xx}(L)$ ($L = 2(n-k)$ is the number of sites in the string) is more intriguing, and the numerical results are shown in Fig. \ref{Fig15}. The inset of Fig. \ref{Fig15} shows that the $O^{xx}(L)$ converges very rapidly with $L$ (except for points in the vicinity of second-order QPT line $J_2=1$). The converged $O^{xx}(\infty)$ monotonously decreases with enhancing the parameter $J_1$ (and hence decreasing $J_2$), and changes continuously through the first-order QPT line $J_1=0$, vanishing immediately after crossing the second-order QPT line $J_2=1$. The asymptotic behavior of $O^{xx}$ near line $J_2=1$ can be predicted by the dual spin correlation function,\cite{Pfeuty,Feng} as $|O^{xx}(\infty)| \sim (1-J_2^2)^{1/4}$, which can be well verified from the fitting in Fig. \ref{Fig15}.

Therefore, the above investigations uncover that in regions I and
II, the string order parameters $O^{xx}$ and $O^{zz}$ are nonzero,
which reveals the hidden $Z_2 \times Z_2$ symmetry breaking in the
EQCM system. In addition, $O^{zz}$ can be used to distinguish two
disordered phases and detect the first-order QPTs between them,
while $O^{xx}$ changes continuously through the transition line
$J_1=0$, and vanishes at critical line $J_2=1$. On the other hand, it is
reported in Ref. \onlinecite{Motamedifar} that the nonzero string
order $O^{xx}$ in disordered region is robust even under some
finite external magnetic fields $h < h_c$ (below the critical field).

\section*{IV. Specific Heat Curves}

\begin{figure}
\begin{center}
\includegraphics[width=0.9\linewidth]{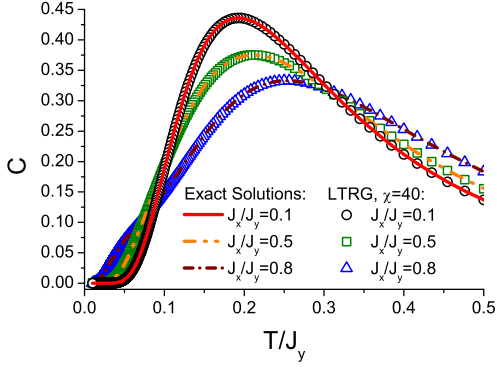}
\end{center}
\caption{(Color online) Specific heat curves for anisotropic XY model. The solid and dashed lines represent the exact solutions, and the scatters are the LTRG results which show perfect agreements with the lines. The couplings $J_x$ and $J_y$ are defined in Eq. (\ref{eq-ani-xy}), $J_y=1$ is set as energy scale here.}
\label{FigS1}
\end{figure}

Besides the ground-state properties, in this section the LTRG method \cite{LTRG} is employed
to investigate the finite temperature properties of 1D EQCM. LTRG
adopts the iTEBD technique for contracting the transfer-matrix
tensor network, and can accurately (and efficiently) obtain the
thermodynamic quantities including free energy, energy,
susceptibility, and specific heat, etc. In Ref. \onlinecite{LTRG},
LTRG method has been applied to calculate the isotropic XY model and achieved
very accurate results. In order to verify the applicability and
accuracy of LTRG for the anisotropic cases (for the present EQCM,
there exist strong anisotropies in spin couplings), the specific heat
curves of anisotropic XY model with Hamiltonian
\begin{equation}
H = \sum_{<i,j>} J_x\sigma_{i}^x \sigma_{j}^x + J_y\sigma_{i}^y \sigma_{j}^y
\label{eq-ani-xy}
\end{equation}
are calculated, and shown in Fig. \ref{FigS1}. The results of LTRG show perfect coincidence with the exact solutions in Ref. \onlinecite{Katsura}.

\begin{figure}
\begin{center}
\includegraphics[width=1.0\linewidth]{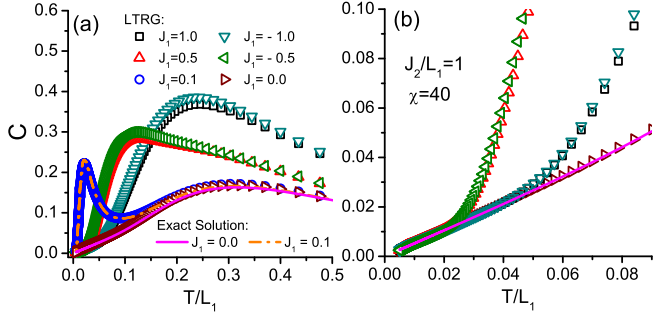}
\end{center}
\caption{(Color online) (a) Specific heat ($C$) curves along the critical line $J_2/L_1=1$. (b) The low temperature sections of $C$. The specific heat curves with the same absolute values $|J_1|$ (but different signs) almost coincide with each other at low temperatures. The exact solutions (see Ref. \onlinecite{Jafari2}) are also plotted with lines, with which the LTRG results show very nice agreements.}
\label{Fig16}
\end{figure}

\begin{figure}
\begin{center}
\includegraphics[width=1.0\linewidth]{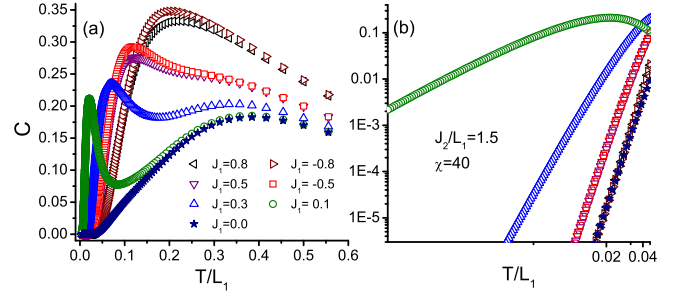}
\end{center}
\caption{(Color online). (a) Specific heat ($C$) curves along the line $J_2/L_1=1.5$. (b) The low temperature parts of $C$, log-log plot reveals the exponential decay explicitly, and the slope of the (nearly) straight part of $C$ is intimately related with the excitation gap. In (b), the lines of $J_1=\pm 0.5$ (as well as $J_1=\pm 0.8$) almost coincide with each other, and $J_1=\pm 0.8$ lines are also very close to the line $J_1=0$.}
\label{Fig17}
\end{figure}

In Figs. \ref{Fig16}, \ref{Fig17}, the specific heat ($C$) curves of 1D
EQCM are present, and close attention is paid to their low-temperature
behaviors, which reveal the low-energy excitation features of
the system. In Fig. \ref{Fig16} (a), the specific heat curves are
evaluated along the critical line $J_2/L_1=1$. It is observed that
with gradually decreasing $|J_1|$, there appear low temperature
sub-peaks moving towards $T=0$, and disappear when $J_1=0$. In Fig.
\ref{Fig16} (b), the low temperature $T$ parts of the $C$ curves are magnified,
and the linear relations with $T$ are clearly shown, which can be
ascribed to the gapless low-energy excitations along the critical
line of EQCM.

Along the parameter line $J_2/L_1=1.5$ , the $C$ curves (versus temperature) are illustrated in Fig. \ref{Fig17}, where sub-peaks also appear and similar movement behaviors are again observed in Fig. \ref{Fig17} (a). It is worth noticing that there exist excitation gaps along the line $J_2/L_1=1.5$, as revealed in Fig. \ref{Fig17} (b), where the low temperature $C$ with $J_1=0$ is shown to decay exponentially. Furthermore, from Fig. \ref{Fig17} (b) (judging from the slop of the $C$ curves in the log-log plot), it is found that the excitation gap $\Delta(J_1)$ tends to zero when the parameters approach $J_1=0$ line from both sides, but the $J_1=0, J_2/L_1=1.5$ point itself is far from gapless.

\section*{V. Summary and Outlook}

\subsection{Subsequent Problems}
The exact soluble 1D EQCM provides an ideal playground for performing calculations and testing MPS-based algorithms, remarkable
accuracy and perfect accordance with previous analytical results have been achieved. Nevertheless, we would like to stress that the
power of MPS-based numerical methods is their accessibility to more complex problems which do not permite exact solutions. Among others, we regard the
following two extended compass models particularly interesting,
\begin{equation}
H_{\rm{XYZ1}} = \sum_{n=1}^{N'} J_{xx} \sigma_{2n-1}^x \sigma_{2n}^x + J_{yy} \sigma_{2n-1}^y \sigma_{2n}^y + J_{zz} \sigma_{2n}^z \sigma_{2n+1}^z,
\end{equation}
and
\begin{equation}
H_{\rm{XYZ2}} = \sum_{n=1}^{N^{\rm{''}}} J_{xx} \sigma_{3n-2}^x \sigma_{3n-1}^x + J_{yy} \sigma_{3n-1}^y \sigma_{3n}^y + J_{zz} \sigma_{3n}^z \sigma_{3n+1}^z,
\end{equation}
where $N' = N/2$, and $N''=N/3$, $N$ is the total site number.
The first one is two-period, with $J_{xx}$ and $J_{yy}$ couplings on
odd bonds and $J_{zz}$ couplings on even ones, the second model is
three-period, with $J_{xx}$, $J_{yy}$, and $J_{zz}$ on three
different types of bonds, respectively. These two models are more
complex than EQCM in Eq. (\ref{Hamiltonian}), while are still very
basic ones. Owing to the existence of three non-commuting spin coupling
components in the Hamiltonian, they can not be diagonalized by
simply taking fermionic transformations, and are expected to show
more interesting QPTs in their phase diagrams. In the first model
$H_{\rm{XYZ1}}$, compared with Hamiltonian Eq. (\ref{Hamiltonian}), the
$J_{zz}$ couplings on odd bonds are replaced with $J_{yy}$
couplings, so the expectation values of parity operators
$\sigma_{2i-1}^z \sigma_{2i}^z$ on odd bonds are no
longer good quantum numbers. Some preliminary results are obtained by
iTEBD calculations, which reveal that there also exist first- and
second-order QPTs, as well as multi-critical points in the phase
diagram of model $H_{\rm{XYZ1}}$. A distinct difference between the phase
diagram of $H_{\rm{XYZ1}}$ and Fig. \ref{Fig1} is that the N\'{e}el and stripe
AF zones in the present EQCM are extending along the $J_1$ axis to the infinity,
while for the former case $H_{\rm{XYZ1}}$, they are confined in a finite region.
More details about the ground-state phase diagrams and QPTs in these two EQCMs will appear elsewhere.

\subsection{Conclusions}

Employing MPS wavefunction, and with the aid of the related
algorithms iTEBD and LTRG, we investigated the ground-state
properties and QPTs, as well as specific heat curves, in the 1D EQCM.

Our calculations, including energy per site, bond energy, entanglement entropy and local magnetizations, validate the phase diagram proposed by previous works. Four different phases are identified in Fig. \ref{Fig1}, including two disordered regions I and II, N\'{e}el ordered phase (region IV), and a stripe AF phase in region III.

The second-order QPTs along $J_2/L_1=1$ line can be detected by the
singularities of entanglement entropy, as well as the derivatives of
bond energy. The first-order QPTs along $J_1 = 0$ are however indeed
missed by entanglement measures according to our calculations.
Furthermore, at the multicritical point ($J_1=0, J_2/L_1=1.0$),
besides the second-order QPT feature revealed by entanglement
entropy, a distinctive ground-state energy level crossing (observed
by taking adiabatic continuations) occurs. Therefore, at the
multicritical point, there coexist both the first-order and the
second-order QPT characters. Furthermore, a logarithmic divergent
behavior of block entanglement $S_{\rm L}$ on the second-order QPT
line $J_2/L_1=1$ are observed, from which the central charge $c=1/2$
is determined.

Fidelity per unit cell is also used to investigate the QPTs, and it is disclosed that both the first- and second-order QPTs in the  EQCM can be detected by identifying the discontinuous and bifurcation points in calculated fidelity curves.

Moreover, the disordered regions I and II are found to possess doubly degenerate entanglement spectra, as well as two types of nonzero string order parameters $O^{xx}$ and $O^{zz}$. By taking dual transformations, it is revealed that the string order parameters reflect the hidden $Z_2 \times Z_2$ symmetry breaking, and parameters $O^{zz}$ can be used to detect the first-order QPT between regions I and II.

Subsequently, the specific heat curves have been studied via LTRG calculations, and low temperature linear behaviors are observed along the critical line $J_2/L_1=1$, while for $J_2\neq L_1$, the exponential decay of $C$ at low temperatures implies the existence of a nonzero excitation gap.

In conclusion, the fidelity per unit cell is shown to be sensitive to detecting not only the first-order but also the second-order QPTs, while the entanglement measures can only detect the latter ones. In the phase diagram Fig. \ref{Fig1}, there exist two symmetry broken phases in regions III (stripe AF) and IV (N\'{e}el) with different local order parameters, and two hidden symmetry broken phases in regions I and II with nonzero string order parameters.

\acknowledgments
The authors would like to thank T. Xiang, J. Vidal, J.H.H. Perk, and Shou-Shu Gong for stimulating discussions, and Xin Yan, Yang Zhao, and Shi-Ju Ran for helpful assistance. This work is supported by the Chinese National Science Foundation under Grant Nos. 11047160, 10874003, and 11004144. It is also partially supported by the National Basic Research Program of China under Grant No. 2009CB939901.

%\begin{thebibliography}{99}

%\end{thebibliography}


\begin{references}
\bibitem{lowest} There exist four eigenstates on odd bond, i.e., $\phi_f$, $\phi_{af}$, $\psi_{f}$, and $\psi_{af}$. For $J_1>0(<0)$ and $L_1=0$, the ground state is $\phi_{f}$ ($\phi_{af}$), and the lowest excited eigenstate is $\psi_{f}$ ($\psi_{af}$), the other two are excited states with higher energies (we constraint the perturbation discussions within the regions $|J_1| \gg |J_2|$). According to the analyses in the text, perturbation within the one-particle subspace (with one odd bond in the excited states) does not modify the ground state up to the first-order approximation.
\bibitem{explan-Ent} The entanglement entropy is continuous in the sence that $S_{\rm{vN}}$ approaches the same value from the left ($J_1<0$) and right ($J_1>0$) sides of the first-order QPT line $J_1=0$. The entanglement entropy just at the first-order transtion point is difficult to determine with the projection algorithm owing to the ground-state degeneracies. Alternatively, we take adiabatic continuation from either side of the line to calculate the entanglement entropies at and across the first-order QPT point, and the resutls also show no singularities.
\bibitem{explan-OP} As mentioned above, the nearest neighbor correlator $\langle \sigma^z_{2i-1} \sigma_{2i}^z \rangle$ can be used to distinguish regions I and II (as shown in Fig. \ref{Fig4}), however, it does not satisfy the definition of local order parameter in the conventional Landau-Ginzburg-Wilson paradigm. Which is more, this corrlator has itimate relation with the non-local string order parameter $O^{zz}$ in Sec. \ref{SOP}.
\bibitem{explan-four-period} As a matter of fact, in order to describe the stripe AF states in region III, four-period MPS should be used in simulations. Therefore, in principle, there exist four half-chain entanglement entropies $S_{4n-3, 4n-2}$, $S_{4n-2, 4n-1}$, $S_{4n-1, 4n}$, and $S_{4n, 4n+1}$ ($n$ labels the four-site unit cell). On the other hand, in practical simultations, all the entropies on the odd (even) bonds are the same up to the calculation errors. Therefore, only two half-chain entanglement entropies $S_{2i-1, 2i}$ and $S_{2i, 2i+1}$ are shown in Fig. \ref{Fig8}.
\end{references}
\end{document}